\documentclass[twocolumn]{aastex62}

\graphicspath{{./}{figures/}}

\received{---}
\revised{---}
\accepted{---}
\submitjournal{ApJ}

\shorttitle{Probing the cosmic ray density in the inner Galaxy}
\shortauthors{Peron et al.}

\usepackage{amsmath}

\begin{document}

\title{Probing the Cosmic Ray density in the inner Galaxy }

\correspondingauthor{Giada Peron}
\email{giada.peron@mpi-hd.mpg.de}

\author{Giada Peron}
\affil{Max  Planck  Institute  for  Nuclear  Physics,  Saupfercheckweg  1,  69117  Heidelberg, Germany}
\author{Felix Aharonian}
\affil{Max  Planck  Institute  for  Nuclear  Physics,  Saupfercheckweg  1,  69117  Heidelberg, Germany}
\affiliation{Dublin Institute for Advanced Studies, 31 Fitzwilliam Place, Dublin 2, Ireland}
\author{Sabrina Casanova}
\affiliation{Max  Planck  Institute  for  Nuclear  Physics,  Saupfercheckweg  1,  69117  Heidelberg, Germany}
\affiliation{Institute of Nuclear Physics PAN, Radzikowskiego 152, 31-342 Kraków, Poland}
\author{Ruizhi Yang}
\affiliation{CAS Key Laboratory for Research in Galaxies and Cosmology, Department of Astronomy, University of Science and Technology of China, Hefei, Anhui 230026, China}
\affiliation{School of Astronomy and Space Science, University of Science and Technology of China, Hefei, Anhui 230026, China}
\author{Roberta Zanin}
\affiliation{Max  Planck  Institute  for  Nuclear  Physics,  Saupfercheckweg  1,  69117  Heidelberg, Germany} \affiliation{CTA observatory, Via Piero Gobetti 93/3
40129 Bologna, Italy }

\begin{abstract}

{ The galactic diffuse $\gamma$-ray emission, as seen by Fermi-LAT, shows a sharp peak in the region around 4 kpc from the Galactic Center, which can be interpreted either as due to an enhanced density of cosmic-ray accelerators or to a modification of the particle diffusion in that region. Observations of $\gamma$-rays originating in molecular clouds are a unique tool to infer the cosmic-ray density point by point, in distant regions of the Galaxy. We here report the analysis of 11 years Fermi-LAT data, obtained in the direction of nine molecular clouds located in the 1.5--4.5 kpc region. The cosmic ray density measured at the locations of these clouds is compatible with the locally measured one.  We demonstrate that the cosmic ray density gradient inferred from the diffuse gamma-ray emission is the result of the presence of cosmic ray accelerators rather than a global change of the sea of Galactic cosmic rays due to their propagation.}

\end{abstract}


\section{Introduction} \label{sec:intro}
Cosmic rays { (CRs)} of energy $\lesssim 10^{15}$ eV are believed to originate inside the Galaxy and to be  confined for at least $\tau \sim 10^{7}$ yr. { During this time, CRs, driven by the interactions with the interstellar medium (ISM) and magnetic fields, mix and spread in the entire Galaxy forming the so-called "Sea" of galactic CRs. The interaction} with the matter of the { ISM produces} detectable $\gamma$-rays  {which tell about the spectral and spatial distribution of the parent  CRs.}(see e.g. \cite{Strong2007}). { At GeV energies, the main contribution to the diffuse $\gamma$-ray comes from pion-decay, resulting from proton-proton interaction \citep{Aharonian2000a}. The outcoming flux is proportional to the gas column $N_{col}$ and CR density, $\rho_{CR}\equiv dN/(dV~dE)$, as:  }
\begin{align}
 \label{eq:fluxgamma} F_\gamma &= \frac{M}{m_p d^2} \frac{c}{4\pi} \xi_N \int dE_p \frac{d\sigma_{pp\rightarrow 2\gamma}}{dE_\gamma} \rho_{CR}(E_p)  \\ & = N_{col} \theta \varphi_\gamma(E_\gamma)\\
 & = \frac{10^5 M_\odot}{m_p~(1~\mathrm{kpc}^2)} A   \varphi_\gamma(E_\gamma)
\end{align}
{ where $M$, $d$ and $\theta$  are the mass, the distance and the angular size of the targeted gas; $\xi_N$ is the nuclear enhancement factor (later on assumed to be 1.8 \citep{Mori2009,Kafexhiu2014}) that accounts for the fraction of heavier nuclei both in the CRs and in the ISM; $d\sigma/dE$ is the differential cross-section of the process \citep{Kafexhiu2014}. In the latter expressions,  $\varphi$ is the $\gamma$-ray emissivity per H-atom, while the number density of the gas can be expressed either in terms of $N_{col}$, or in terms of the factor $A\equiv M_5/d^2_{kpc}$ , where $M_5=M/10^5~M_\odot$ and $d_{kpc}=d/1~kpc$ particularly useful when dealing with molecular clouds. }

Several studies have been dedicated to analyze the galactic diffuse $\gamma$-ray emission from the pioneering studies with COS-B \citep{Strong1988} and EGRET  \citep{Strong1996} to the more recent investigation with the Fermi Large Area Telescope (LAT) \citep{Acero2016,Yang2016, Pothast_2018}. These works { (hereafter referred as {\it ring analyses})} divide the ISM in galactocentric rings and extract the $\gamma$-ray emissivity in each of them. In all cases the authors report a hardening and an enhancement of the emissivity { in} the inner galactocentric regions, with a maximum at a distance of $\sim$ 4 kpc from the Galactic Center (GC). \cite{Acero2016} reported the highest enhancement ($\sim$ 400\%) and the hardest spectrum in the 1.5--4.5 kpc region, \cite{Yang2016} found similar results in the 4--6 kpc ring. \cite{Pothast_2018} found the largest enhancement { in the region } 1.7--4.5 kpc and the hardest spectrum { within 4.5 kpc and 5.5 kpc  from the GC}.  On the other hand, the intensity measured in the innermost part of the Galaxy ($<$ 1 kpc) is significantly lower, with values comparable to the local CR flux \citep{Acero2016,Pothast_2018,Yang2016}. A similar density  is also observed in the molecular clouds of the Sagittarius B Complex at the very center ($\sim$ 100 pc) of the Milky Way \citep{Yang2015,Aharonian2020}. { Ring analyses of the diffuse $\gamma$-ray emission can be biased by the presence of a few regions with enhanced CR density. Furthermore, the diffuse $\gamma$-ray emission measured in galacto-centric rings is artificially assumed to be cylindrical symmetric. In this way the resulting CR density is an average on a very large area and no fluctuation within the ring can be detected.    }

Molecular clouds (MCs), being isolated dense clumps of the ISM, provide unique conditions for testing the CR density in different regions of the Galactic Disk. In fact, while studies of the diffuse emission can only provide {\it integral} information, MCs give {\it differential} information. Namely, they yield the values of the CR spectrum at their location, { making it possible} to trace the distribution of galactic CRs point by point \citep{Casanova2010}. This solves any bias due to the symmetry hypothesis introduced in the ring analyses. Earlier we reported \citep{Aharonian2020} the $\gamma$-ray spectra from Fermi-LAT observations of a dozen of giant MCs from the catalog of  \cite{Rice2016a}.  { The analysis was based on CO and HI templates that permitted to trace the three dimensional distribution of the gas and therefore to extract the spectrum just from the region of the cloud. Any effect of contamination from the gas on the line of sight (l.o.s.), was taken under control by the selection of clouds which dominated the column in terms of gas column density.} Our results demonstrated for the first time the feasibility of this method and revealed possible fluctuations of CR densities for clouds located at similar  { galacto-centric} distances, in some cases { matching} the same level as measured in the vicinity of { the} Earth.  We argued { for} a uniform sea scenario with the CR density enhanced only in some locations, corresponding to regions with a higher density of accelerators. However, the large systematic uncertainties prevented us to draw a robust conclusion. Moreover, the considered catalog of MCs \citep{Rice2016a} excluded the innermost regions ($|l|<13^{\circ}$) of the Galaxy, { so that we could not extract information} from the central ($<$ 4 kpc) part. 

Here we present the analysis of the { $\gamma$-ray emission of} the gas column in the direction of nine GMCs from the catalog of \cite{Miville-Deschenes2016}, located within 1.5 and 4.5 kpc. This region is of special interest both because it was unexplored by previous studies and also because it is expected to present the highest and hardest $\gamma$-ray emissivity, { according to the ring-based analysis presented in \cite{Acero2016}}. Such enhancement facilitates the detection of MCs. { Moreover, it is interesting to test, whether the higher CR-density characterizes the whole ring, or if it is a result of summing localized regions of enhanced CRs}. The detection of fluxes under-luminous with respect to the value reported { for the corresponding ring}, would pose severe constraint on the nature of the enhancement and the hardening in this zone.

\section{SELECTED TARGET MOLECULAR CLOUDS}
The recently released catalog of \cite{Miville-Deschenes2016}, hereafter referred as MD-catalog contains 8107 object and is by far the most comprehensive catalog of { MCs. The latter} covers 98\% of the molecular medium traced by \cite{Dame2000}, including as well the inner galactic region, that was unexplored in other catalogs, { such as in} \cite{Rice2016a}.  The MD-catalog spans the entire galactic disk { over}  the latitude range $|b|<5^{\circ}$, and includes more than 300 MCs with $M>10^6$ M$_{\odot}$, of which 25\% resides between 1.5 kpc and 4.5 kpc { from the Galactic Center}.

The distance of a MC is the measurable that suffers of the largest uncertainties, especially in the innermost part of the Galaxy. The galactocentric distance to each molecular cloud, $R_{gal}$, is typically assigned via the {\it kinematic distance method} \citep{Roman-Duval2009}, which relates the observed radial velocity, $v_{LSR}$, to the rotation velocity of the Galaxy. However, because of the dependency of this relation on $\sin(l)$, at low longitudes ($l \lesssim 10$) the typical broadening in velocity, $\sigma_v  \sim$ 10 km/s, results in a big difference in distance. Therefore an accurate kinematic separation of the diffuse gas components { along the l.o.s.} cannot be realized at these longitudes. In the case of clouds, the accuracy on the kinematic distance can be improved by cross-correlating the cloud coordinates ($l,b,v$) with the coordinates of spiral arms or other objects with precise parallax determination, as described in \cite{Reid2016}. For the clouds of interest, we checked the distance with this method,  using the online available tool \footnote{http://bessel.vlbi-astrometry.org/bayesian} and have been convinced that they were compatible with the values reported in the catalog. This is of { paramount} importance  { to } us, since our main interest is indeed to observe Molecular Clouds in the inner regions of the Galaxy, specifically in the 1.5--4.5 kpc ring. 

This region is particularly complicated to analyze also because, towards the Galactic center, several spiral arms overlap and consequently many sources lay on the same l.o.s.. { In order to mitigate source confusion}, as a first selection criterion, for our analysis, we discarded those clouds which overlap with any known 4FGL source \citep{TheFermi-LATcollaboration2019}. Secondly, we { had to guarantee that the observed diffuse emission originated in the 1.5--4.5 kpc region and not from gas that interposes along the l.o.s.. Thus, we} chose those clouds that give a major contribution to the l.o.s. both in terms of gas density and in terms of possible enhancement of the gamma-ray emission. { In this way, in the observed direction, most of the emission is expected to originate in the 1.5--4.5 kpc region, and a minor fraction, from the rest of the column, that does not belong to the ring. Since we are interested to see if the flux in the given region is more similar to the enhanced value reported by \cite{Acero2016} or to the local spectrum,} we estimated the { maximum fraction of column density, $X$, that could be outside of the  1.5--4.5 kpc region } in order to  be able to distinguish between a local and an enhanced CR flux. If $N$ is the enhancement in gamma-ray flux with respect to it in the remaining $(1-X)$ part of the column, the maximum fraction of gas that we can have in the background is calculated, from the condition:

\begin{equation}\label{eq:cond}
    (1-X) \cdot N \cdot  \ F_{bg} - X  \cdot F_{bg} >  0.3 [(1-X)\cdot  N\cdot F_{bg} +X\cdot F_{bg}  ].
\end{equation}

Equation \ref{eq:cond} implies that the measured flux from the column could be distinguished, if enhanced from a background-type flux with a separation of at least 30\%. That is the typical level of uncertainty of the gas column density.  For example we can easily see that if the $\gamma$-ray emissivity is enhanced by a factor $N=4$ in the region of the cloud, we could measure such enhancement over the background as far as $X<68 \%$.  { The fraction $X$ can be calculated from CO maps (\cite{Dame2000}). That allows a first screening, but it is not free from the uncertainties due to the gas kinematic separation. Furthermore, the { l.o.s.} might include gas from different regions of the Galaxy with different emissivities, intermediate between the local and the 1.5-4.5 kpc one. So, for the firstly-selected clouds the expected flux from the entire l.o.s. has been estimated in two cases and checked that they were distinguishable of an adequate level (see Fig. \ref{fig:sed_all}). A {\it uniform scenario}, where the gas along the l.o.s. emits with a constant emissivity, similar to the local value, $F^{}_\gamma(E) \propto A_{tot} \times F^{loc}(E)$ (orange curve in Fig. \ref{fig:sed_all}), and a {\it radial dependent} scenario, where the emissivity depends on the galacto-centric location, $r$ , of the gas, $F_\gamma(E) \propto \sum_{r_i} A_{r_i} \times F^{r_i}(E_p,r_i)$ (blue curve in Fig. \ref{fig:sed_all}) have been considered. For the latter case, we considered the gamma-ray fluxes, $F^{r_i}$, of the rings derived by the Fermi-LAT collaboration. The local value, $F^{loc}$, has been assumed to coincide with the value measured in diffuse analyses at the 8--10 kpc ring. Note that in principle this should not necessarily be the same as the value measured by direct experiments in the vicinity of Earth, e.g. AMS02 \citep{Aguilar2015}. The red curve in Fig. \ref{fig:sed_all}, shows the flux derived from a PL spectrum of index 2.8, normalized at the value measured by AMS02 at 100 GeV. The latter is slightly steeper than the flux of the ring, resulting in a factor of 2 lower at energies above a few GeV. The expected fluxes calculated for each specific ring are also shown in Fig. \ref{fig:sed_all}. The expected contribution of the 1.5--4.5 kpc ring (cyan dashed line) is dominant for the selected regions. 

To evaluate the fraction of gas density that falls in each ring, we determined which clouds of the MD-clouds overlapped (also partially) with the area of interest. Clouds have a better determined distance with respect to the diffuse gas. In the case of partial overlapping, we considered a fraction of mass, that corresponded to the fraction of pixels that fell in the considered area. Then, knowing the galctocentric distance of the clouds from the catalog, the mass can be easily partitioned in the galactocentric rings. Because of the high degree of completeness of the catalog (98\%), molecular clouds of the MD-catalog trace completely the molecular medium. The correspondence between clouds and diffuse is also tested by the authors of the catalog \cite{Miville-Deschenes2016} by comparing the values of surface density derived from clouds, to those derived with diffuse and assessed a good agreement between the two quantities. As a cross-check, for the selected regions, the $A$ parameter derived from the dust column density has been compared to the one derived as a sum of the ratio $M_5$/d$_{kpc}^2$ of each cloud. The two estimations gave comparable results (see Table \ref{tab:clouds}). Small difference can arise both because in the case of clouds the mass is considered to be uniformly divided among all the pixels, and this is often not the case, and because the CO and dust template might differ in some locations. }

\begin{table*}
\centering

\begin{tabular}{lcccccccccc}
\hline \hline
\#  & ($l,b$)  & $v$ & $\theta$ & $d$ & $d^{px}~ (\mathcal{P})$&  $R_{gal}$ & $A^{dust}_{tot}$& $A^{MC}_{tot}$   & $X$ \\
  & [$^\circ$] & [km s$^{-1}$] &   [$^\circ$]   & [kpc]& [kpc] & [kpc] & & & \\
\hline
57  &  (2.21, -0.21) &8.43    & 0.55 & 12.69 & 13.64(0.4)  & 4.21 & 2.59 & 2.68 &   0.47 \\
78  & (2.93,  0.27)  &25.85   & 0.33 & 10.94 & 10.81(0.6)  & 2.49 & 1.07 &1.83 &  0.63 \\
120 &  (22.46,0.16)  &89.33   & 0.31 & 10.3  &  9.57(0.7)  & 4.06 & 0.77 & 0.78 &    0.34 \\
135 & (24.4, -0.09)  & 112.09 & 0.33 & 6.45  & 6.06(1.0)   & 3.74 & 0.91 &1.06 & 0.46 \\
148 & (342.2, 0.26)  & -79.0  & 0.46 & 5.26  & 5.27(0.9)   & 3.85 & 1.47 & 1.35 &  0.47 \\
368 & (5.43, -0.38)  & 20.87  & 0.28 & 12.57 & 12.38(0.75) & 4.19 &  0.49 & 1.11 &   0.50 \\
411 &(23.71,  0.31)  & 108.99 & 0.35 & 6.26  & 6.03 (0.9)  & 3.74 & 1.04 & 0.60  &    0.28 \\
1155& (3.93, -1.02)  & 59.32  & 0.47 & 10.01 & 8.92(0.7)   & 1.64 & 0.67 & 0.85 &    0.35 \\
1312&  (351.5, 0.22) & -43.21 & 0.46 & 11.76 & 11.47(0.6)  & 3.58 & 1.11 & 1.08 &  0.43 \\

\hline \hline
\end{tabular}
\caption{Parameters of the selected l.o.s.. The numeration and the physical parameters ($l,b,v, d$ and $R_{gal}$) are taken from the MD-catalog. The distance derived with the parallax-calibrated rotation curve is also shown ($d^{px}$),together with the value of probability, $\mathcal{P}$, that the cloud falls in that location. The total $A$ parameters are calculated from dust and from the sum of clouds as explained in the text. $X$ is the fraction of gas that does not belong to the 1.5--4.5 kpc ring.}
\label{tab:clouds}
\end{table*}

\section{OBSERVATIONS}
We used \texttt{fermipy v.0.17.4} to analyse Fermi-LAT \texttt{PASS8} data accumulated for more than 10 years, from August 4th 2008  (\texttt{MET 239557417})  to January 8th 2020 (\texttt{MET 600181346}) in the direction of the chosen MCs. We selected \texttt{FRONT\&BACK} events and imposed \texttt{DATA\_QUAL==1 \&\& LAT\_CONFIG==1}.  To minimize the contribution from the Earth limb we considered only events with zenith angle smaller than 90$^{\circ}$.  In the starting model we included the sources from the 4th Fermi-LAT  (4FGL) Source Catalog \citep{TheFermi-LATcollaboration2019}.  { Only photons of energy $>$ 1 GeV have been considered, to benifit of the better angular resolution.}

\begin{figure*}
\includegraphics[width=0.9 \linewidth]{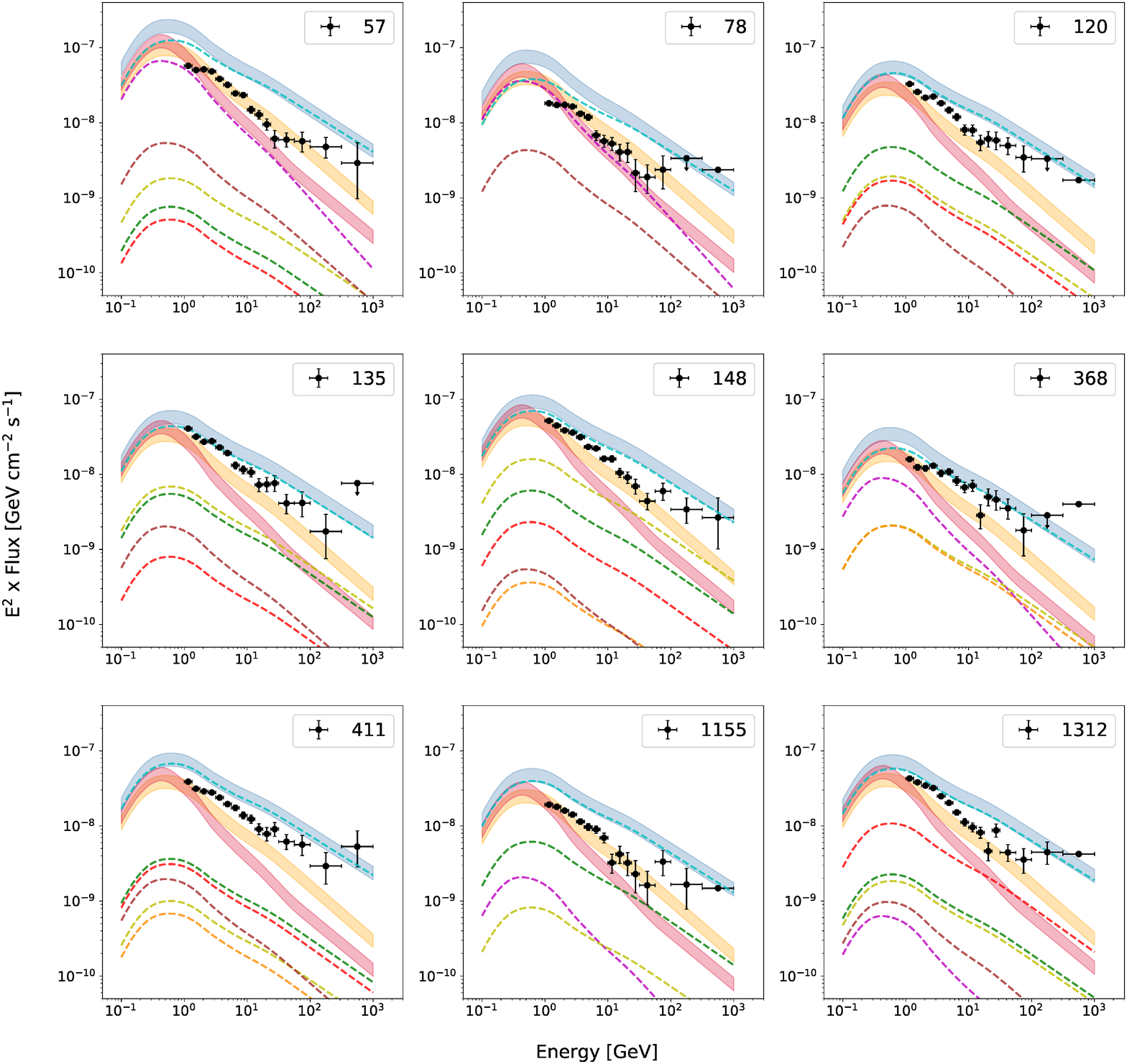}
\caption{The SED derived from the direction of the selected MCs. The spectral points compared with the theoretical flux calculated for a uniform scenario (thick orange area) and for a radial dependent scenario (thick blue area). The areas represent the 20\% uncertainty in the mass that derives from the used tracer, dust. The contribution of each ring to the los, $F_{r_i}$, is also plotted, respectively: 0--1.5 kpc (magenta), 1.5--4.5 kpc (cyan), 4.5--5.5 kpc (green), 5.5--6.5 kpc (yellow), 6.5--7 kpc (orange), 7--8 kpc (red), 8--10 kpc (brown), 10--16.5 kpc (violet), 16.5--50 kpc (blue).}
\label{fig:sed_all}
\end{figure*}

Following the same { methods as in \cite{Yang2015} and \cite{Aharonian2020}, for the cases of Gould Belt clouds and Sgr B, we analyzed the entire column of gas in the direction of the selected clouds. The selection procedure assures that the emission originating from the 1.5--4.5 kpc region is the dominant component. Moreover, in the considered region, a kinematic separation of the gas would have not been reliable for the arguments presented above. } We constructed a customized  model for the galactic { diffuse} background emission that includes the Inverse Compton, produced by \texttt{galprop} \citep{Vladimirov2011}, and a spatial template for pion emission based on Planck dust map. Specifically, we considered, as a tracer of interstellar gas, the dust opacity map at 353 Hz that shows a linear relation with the gas column density \citep{Ade2011}. The advantage of dust optical depth, is that it traces both molecular and atomic hydrogen and it is not subject to saturation, hence it traces also the so-called {\it dark} gas. { This is of particular importance when observing inner galactic regions, since at high column densities CO easily saturates. }  Besides, dust is not subject to the uncertainty on the parameters that characterize the CO and HI emission, namely the conversion factor and the spin temperature.  For each MC, we created then a background from the dust map { for the entire region of interest (ROI), excluding the portion of gas} centered at the location of the cloud and with the corresponding size. This allows us to analyze the extracted gas as separate component and to extract the spectrum from there. 
We proceed by optimizing the model and by fitting all the sources within 3 degrees from the center and the normalization of all the brightest sources (TS$>$100). After the first fit, we evaluated the Test Statistic (TS) map, included in the model any excess with TS$>$25 and refitted until the residuals became negligible.  We then extracted the spectral energy distribution (SED) for each cloud, by fitting a Power Law of index 2 in each energy bin. 
The SED {derived in the direction of} all the clouds are shown in Fig. \ref{fig:sed_all}, together with the expected fluxes evaluated as explained before. We can see that the SED  not always matches the expected enhanced values. On the contrary, it is often lower. Remarkably, in some cases the measured flux level is the same as the one measured in the local ring. For what concerns the slope, we also see that the spectrum is not always as hard as expected and sometimes it is much softer.

We then interpolated the { spectral point}  with the python package \texttt{naima v.0.8.1} \citep{naima}, { that allows, assuming a given radiative mechanism, the determination of the spectral parameter of the parent particles. In this case, pion decay has been considered as the main contributor to the emission and  the parent proton spectrum has been derived (cf. Eq. \ref{eq:fluxgamma})}. We used a Power Law model to interpolate the CR spectrum: 

\begin{equation}
\rho_{CR}(E) = \rho_{0,CR} \bigg( \frac{E}{30 ~\mathrm{GeV}} \bigg)^{-\alpha_{CR}}
\end{equation}

\begin{figure*}
\includegraphics[width=1 \linewidth]{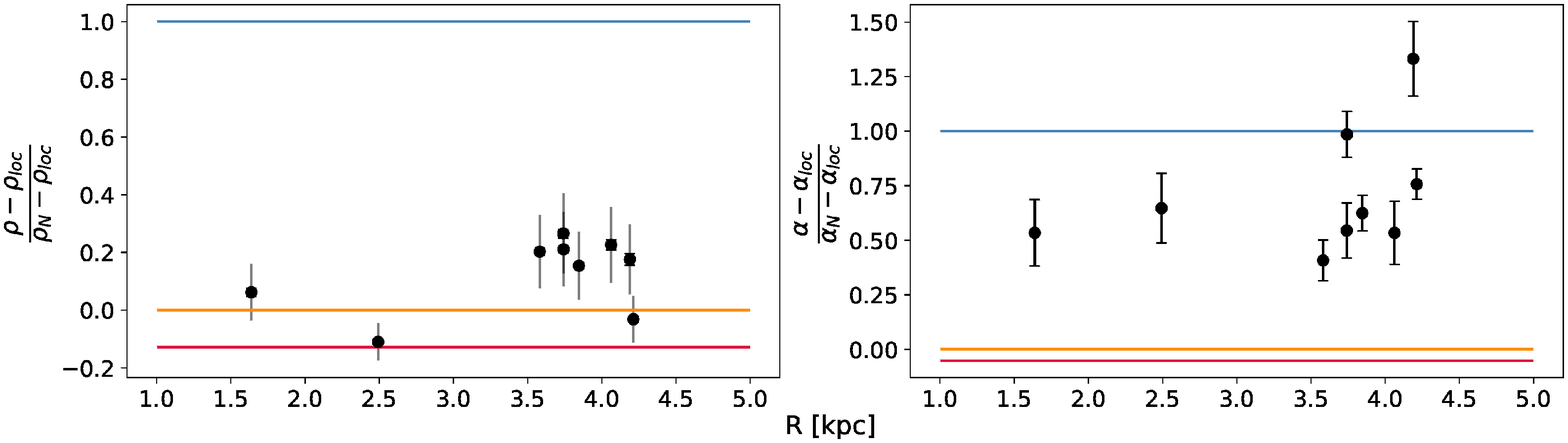}
\caption{Left panel:  ratio between the difference between the CR normalization found in the clouds at 30 GeV, $\rho$ and the CR normalization found in the local 8--10 kpc ring, $\rho_{loc}$, and  the difference between the CR normalization measured in the 1.5--4.5 kpc, $\rho_N$, ring and the local ring. Right panel: same quantity calculated for the index of the CR spectrum. The red line indicates the ratio value calculated for the CR density measured by AMS02, while the orange and blue lines indicate the scenario where the density coincides with the one extracted in the local ring ($\rho=\rho_0$) or to the one extracted in the enhanced ring ($\rho= \rho_N$).  }
\label{fig:radial}
\end{figure*}

We chose to consider a pivot energy of 30 GeV for protons, since the corresponding gamma-ray observations starts from 1 GeV and therefore a higher value of energy is better constrained. 
The results for the normalization and the spectral index are presented in Table \ref{table:naima} together with the corresponding values derived from from the points in \cite{Acero2016} in the rings of interest. The latter have been newly interpolated, starting from the reported emissivities (Fig. 7 of \cite{Acero2016}). The ratios between the difference from the cloud parameters and the local ones, and the parameters of the enhanced ring to the local ring, namely: 
$$\frac{\rho - \rho_{loc}}{\rho_N - \rho_{loc}}$$and $$ \frac{\alpha - \alpha_{loc}}{\alpha_N - \alpha_{loc}},$$are plotted in Fig. \ref{fig:radial}. 
The ratios measure the compatibility of each  cloud to one or the other scenario: if the ratio is close to 0, the uniform local scenario is preferred, whereas if it is closer to 1, the radial dependent scenario wins.  A certain degree of scatter is observed and, compatibly with the considerations on the SEDs, the normalization derived for few clouds is in agreement with the uniform scenario.  { Note that $\rho$ is extracted from the entire column, as if all the gas is condensed in the region of the cloud and for the derived values constitute an upper limit to the CR density that is truly in the 1.5--4.5 kpc region. This constrains even more the points towards the direction of a lower flux. In Table \ref{table:naima} are also reported the values, $\rho*$, of CR density obtained after the subtraction of the fraction of gas, $X$, that belongs to the background, assuming that it contains a value of CR density similar to the local (8-10 kpc ring) value.}

\begin{table}
\centering
\small
\begin{tabular}{lccc}

\hline \hline
& $\rho_{0,CR}$ (30 GeV) & $\rho^*$ (30 GeV) & $\alpha_{CR}$ \\
&  \multicolumn{2}{c}{[$10^{-13}$ GeV$^{-1}$ cm$^{-3}$ ]} & \\
\hline
57 & 0.88 $\pm$ 0.01   & 0.43    & 2.636 $\pm$ 0.014 \\
78 & 0.71 $\pm$ 0.02   & 0.11    & 2.66 $\pm$ 0.03 \\
120 & 1.45 $\pm$ 0.04  & 1.13    & 2.68 $\pm$ 0.03 \\
135 & 1.54 $\pm$ 0.03  & 1.10    & 2.68 $\pm$ 0.03 \\
148 & 1.29 $\pm$ 0.02  & 0.84    & 2.663 $\pm$ 0.016 \\
368 & 1.34 $\pm$ 0.04  & 0.86    & 2.52 $\pm$ 0.03 \\
411 & 1.42 $\pm$ 0.03  & 1.15    & 2.59 $\pm$ 0.02 \\
1155 & 1.09 $\pm$ 0.03 & 0.76    & 2.68 $\pm$ 0.03 \\
1312 & 1.4 $\pm$ 0.02  & 0.99    & 2.707 $\pm$ 0.019 \\
\hline
1.5--4.5 kpc & 3.15 $\pm$ 0.17 &-- &2.587 $\pm$ 0.016 \\
8--10 kpc & 0.95 $\pm$ 0.05 &  --&2.790 $\pm$ 0.017 \\
AMS02 & 0.67 & -- & 2.97 \\
\hline \hline

\end{tabular}
\caption{CR parameters (density and spectral index) derived from the interpolation of the SED with a Pion decay model of emission with naima. $\rho^*$ is the CR density obtained after subtracting the background density, assumed to coincide with the one of the 8--10 kpc ring. The corresponding values for the ring of interest and for a proton flux similar to the one measured by AMS02 are also shown.Note that these values correspond to a total energy reppresentation }
\label{table:naima}
\end{table}

\begin{figure}
\includegraphics[width=1 \linewidth]{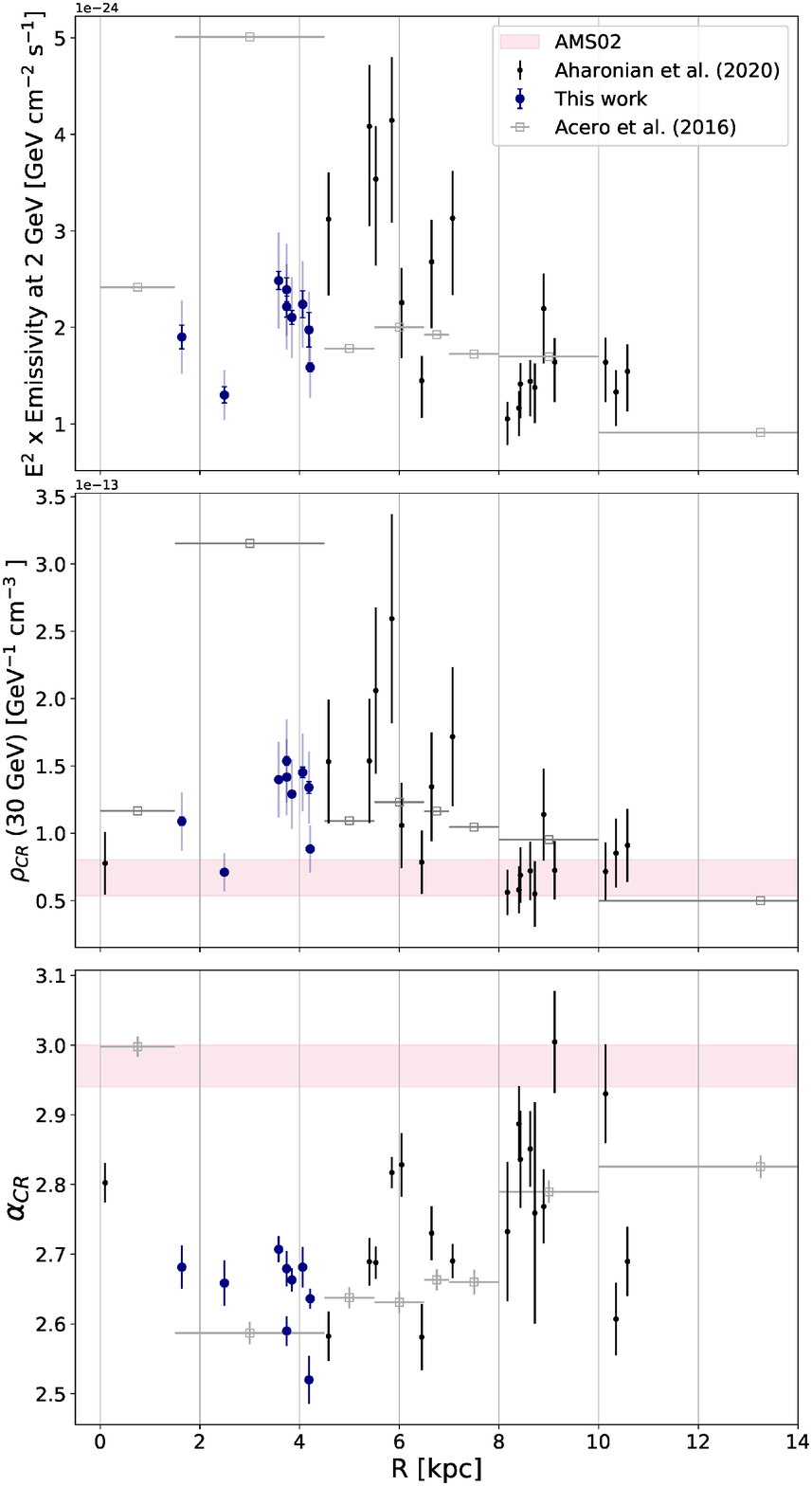}
\caption{The Gamma-ray emissivities and the proton parameters derived for the analyzed clouds (blue bullets) are shown as a function of their galactocentric locations; the light errorbar indicates the systematic uncertainty that derives from the gas column density. The points are compared with the value derived for the clouds analyzed by \cite{Aharonian2020} (black points) and with the value derived from the analysis of the diffuse (grey squares). The red area indicates the values corresponding to a proton flux that coincides with the one measured by AMS02, the extent of the area accounts for the 20\% systematic uncertainty that derives from the gas column density.}
\label{fig:radial_compare}
\end{figure}

 The observed emissivities and the derived CR densities and spectral indexes are compared in Fig. \ref{fig:radial_compare} with the values derived in different rings by \cite{Acero2016} and the values extracted in GMCs by \cite{Aharonian2020}. Both have been re-normalized to the energy of 30 GeV. The values corresponding to the proton spectrum measured directly by AMS02 in the vicinity of Earth are also indicated as red area. 

\section{DISCUSSION }
 
The results derived { here from} MCs in the 1.5--4.5 kpc region depart { from} the values reported in the ring-analyses studies. In this region the analyses of diffuse gas showed enhanced and harder density of CRs, with respect to the local values,  while the values extracted from clouds are much closer to the local spectrum. Several theoretical models have tried to explain the observed behaviour in the specific ring either by introducing a second component of freshly accelerated particles \citep{Guo2014} or by assuming a radial dependent diffusion coefficient \citep{Gaggero2015,Guo2018}. The first well explains the observed diffuse $\gamma$-rays, but can't explain other observed effects, in particular the hardening of the primaries. The second well reproduces almost all the observable parameters, but fails to reproduce the low emissivity detected in the galactic center \citep{Yang2015,Aharonian2020}.  { The contribution of unresolved $\gamma$-ray sources to the observed enhancement have been evaluated by \cite{Pothast_2018} , that found it not sufficient to explain the observation at GeV energies. } Furthermore, in a recent work, \cite{Cataldo2019} showed that the extrapolation to TeV energies of the observed enhanced flux in the inner Galaxy  saturates the observed emission due to pion decay and known point sources, leaving no additional space for unresolved sources and other galactic diffuse emission components. Our results on specific locations inside the ring allow us to pose clear constraints on the above mentioned theories. Indeed the observation of a systematically lower flux with respect to the large-scale value together with the variability of the parameters from source to source {  casts serious doubts on }the possibility of a global variation of the level of the CR sea on kpc scales, modulated by the CR propagation in the galactic magnetic fields. This behaviour rather  { agrees}  with the scenario of a uniform sea of CRs, altered in specific locations { by the} presence of active accelerators.  It is in fact striking that in several cases the upper limits are very close to the flux of the local 8--10 kpc ring and in one case it matches exactly the value measured by AMS02 in the vicinity of Earth. 
{ It it clear that such results could not be obtained in the analyses of large scale diffuse emission, as only a single region with significantly larger flux, compared to the nominal value, if included in the target, would increase the measured average CR density. Note in fact, that the diffuse gas, being dominated in density by giant molecular clouds, suffers of a selection effect, meaning that the measured value is representative of the most dense regions, which are the ones that dominate in mass the observed target. If the content of CRs is enhanced in some or all of those dense regions, also the mean value assigned to the CR density would result altered. This is a reasonable argument that could explain the observed enhancement in the inner galactocentric regions on a large scale. Moreover it would explain the differences that emerge in different authours' work, since choosing different regions to analyze would include different contributions. Also, it agrees with the observations of GMCs, \citep{Aharonian2020}, which fluctuate below and above the average values derived for the corresponding ring (see Fig. \ref{fig:radial_compare}). Finding instead a column of gas, where no enhancement with respect to the local value, is detected is an exceptional case, especially if that column crosses the inner galactocentric regions, where the most massive clouds and active accelerators reside.} 

Note moreover that our method is free from the conventional uncertainties that limit these kind of investigations. In particular, the choice of using dust opacity maps instead of HI+CO allows us to overcome several issues that concern these tracers.  Firstly, the uncertainty on the HI spin temperature, that is hardly determined a priori, and is normally assumed to be constant throughout the entire Galaxy. Secondly the uncertainty on the $X_{CO}$ conversion factor for which it is not clear if a constant value, similar to the local one can be assumed, as metallicity or stellar radiation fields may have a significant influence on it. { The latter in particular significantly influences the interpretation of the results. Note for example that in the analyses of diffuse gas, the gradient of emissivity turns out smaller ($\sim$ 50\%) when using a constant X$_{CO}$ factor as in \cite{Pothast_2018}, although in this case such shallow variation would not match the peaked distribution of progenitors \citep{Green2015}, as pointed out in \cite{Strong2004}. Nevertheless \cite{Yang2016} observe a high enhancement in the inner Galaxy also when using dust as a tracer of the interstellar medium, suggesting that the enhancement does not only originate from the uncertainties of the CO-to-H$_2$ conversion factor. } Dust opacity maps suffers of less uncertainties and has also the advantage of being sensitive to the non-traced (by HI and CO) gas. The correspondence between the column density, derived from dust opacity, and the value derived from clouds (see Sec. 2), guarantees that there is a linear relation between the two tracers in the considered regions. This, together with the completeness of the MD-catalog, also justifies the choice of dividing the column density contribution based on clouds. We are convinced that this is a more appropriate approach, since clouds are identified as peaks and hence are less sensitive to the spread in velocity. Besides, the accuracy of the distance is assured by the cross correlations with objects of known parallax.  

In summary, we extended our previous study  \newline \citep{Aharonian2020} of the CR-density distribution in the Milky Way based on the Fermi LAT gamma-ray observations of individual giant molecular clouds, to the galacto-centric distances 1.5-4.5 kpc. Using the data on the dust component, we reduced the uncertainties in the extraction of the mass of clouds.  We do not confirm the high emissivity reported by \cite{Acero2016} and other authors in the region of 1.5--4.5 kpc, but we rather observe in the direction of every targeted cloud a significantly lower flux. This is an important results not only as it rules out the possibility of a global modification of the sea of CRs, but also because it proves that the results obtained by diffuse analyses might be biased. The new results also support and extend to the 1.5--4.5 kpc region the tendency of fluctuations of the CR spectral parameters earlier reported in \cite{Aharonian2020}.
This behaviour favours the existence of a uniform sea of  galactic CRs with the exception of regions where active or recent acceleration alters its level. The nominal value of the sea must be similar to the one measured at Earth,  as demonstrated by the values measured in several different locations from the outermost ($>$ 10 kpc) to the innermost ($<$ 1 kpc) regions \citep{Aharonian2020}, and now also in the 1.5--4.5 kpc ring.

\software{astropy v.2.0.9 \citep{astropy:2018},  fermipy v.0.17.4 \citep{Wood2017}, naima v.0.8.3 \citep{naima} }

\acknowledgments
R.Z. is supported by the NSFC under grants 11421303 and the national youth thousand talents program in China. S.C. acknowledges Polish Science Centre grant, DEC-2017/27/B/ST9/02272.

\newpage

\appendix

\section{Complements to target selection}
\begin{figure}
    \centering
    \includegraphics[width=0.6\linewidth]{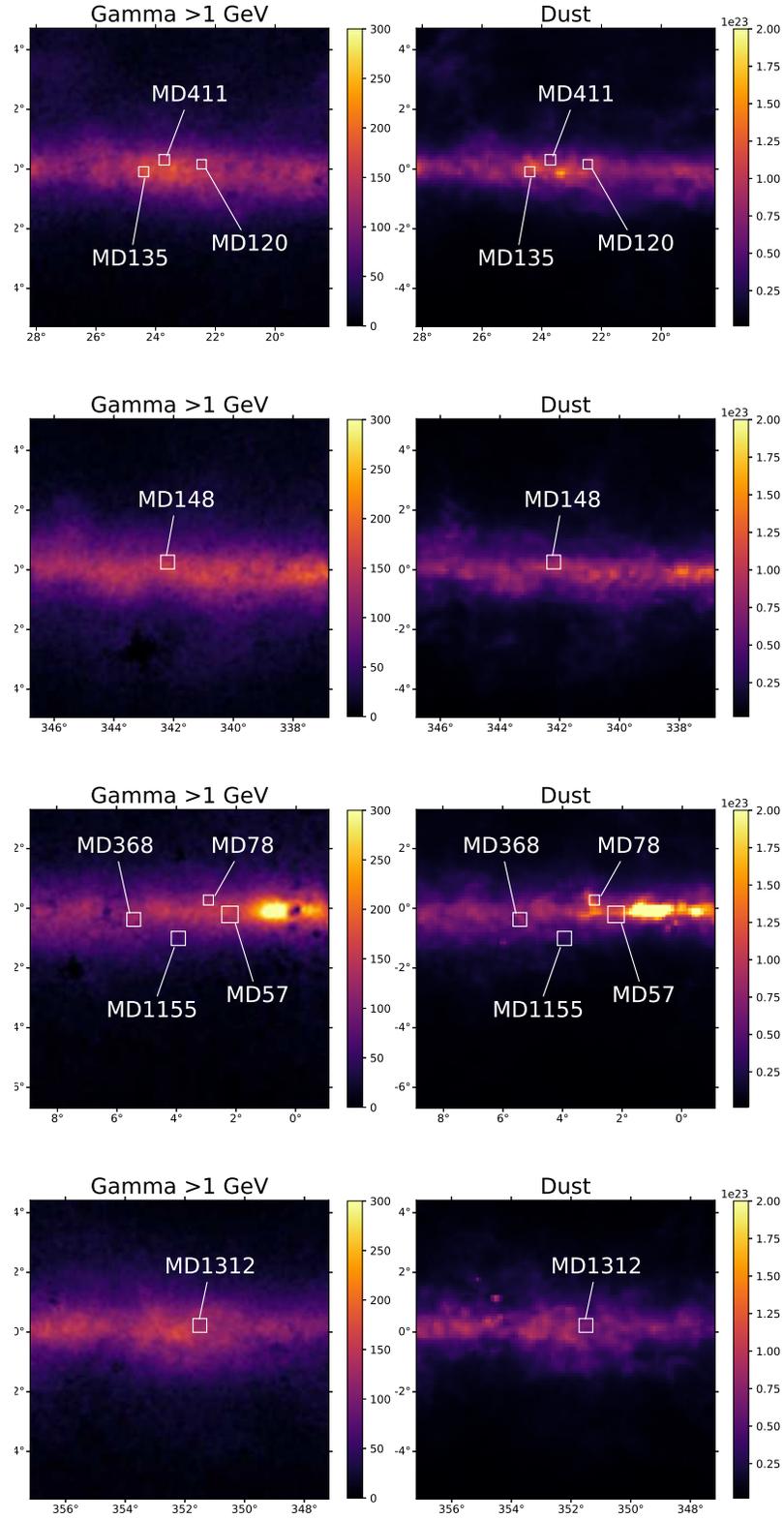}
    \caption{Maps in galactic coordinates ($l,b$) of the gamma-ray counts ($>$ 1 GeV) after the subtraction of the 4FGL sources and of the corresponding dust opacity in the regions of the analyzed molecular clouds.  } \label{fig:gamma-dust}
\end{figure}

\begin{figure}
    \centering
    \includegraphics[width=1\linewidth]{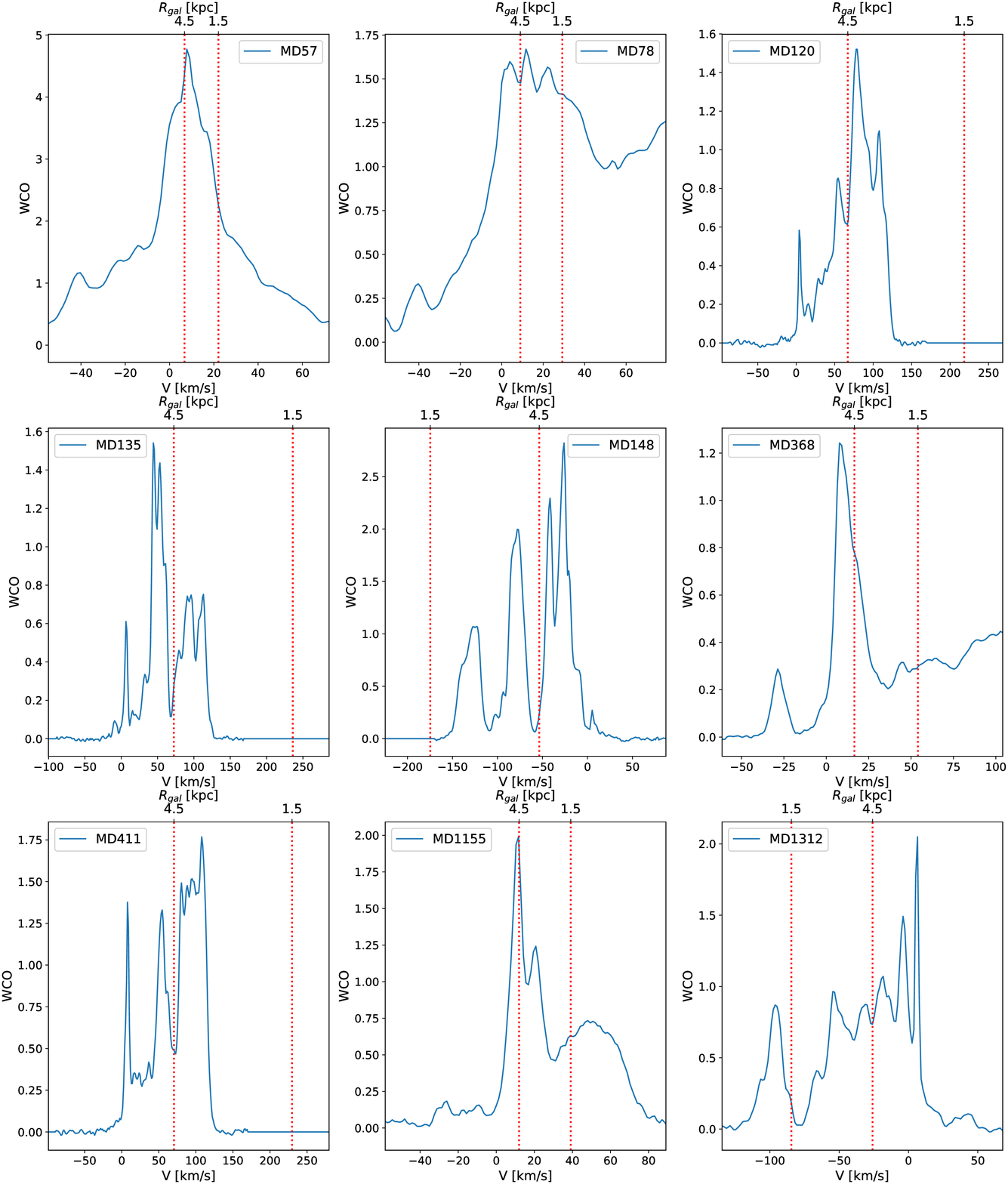}
    \caption{The velocity distribution of the gas, traced by CO, in the analyzed regions. The edges of the galactocentric ring 1.5--4.5 kpc are indicated in red. }
    \label{fig:los-clouds}
\end{figure}

\begin{figure}
    \centering
    \includegraphics[width=1\linewidth]{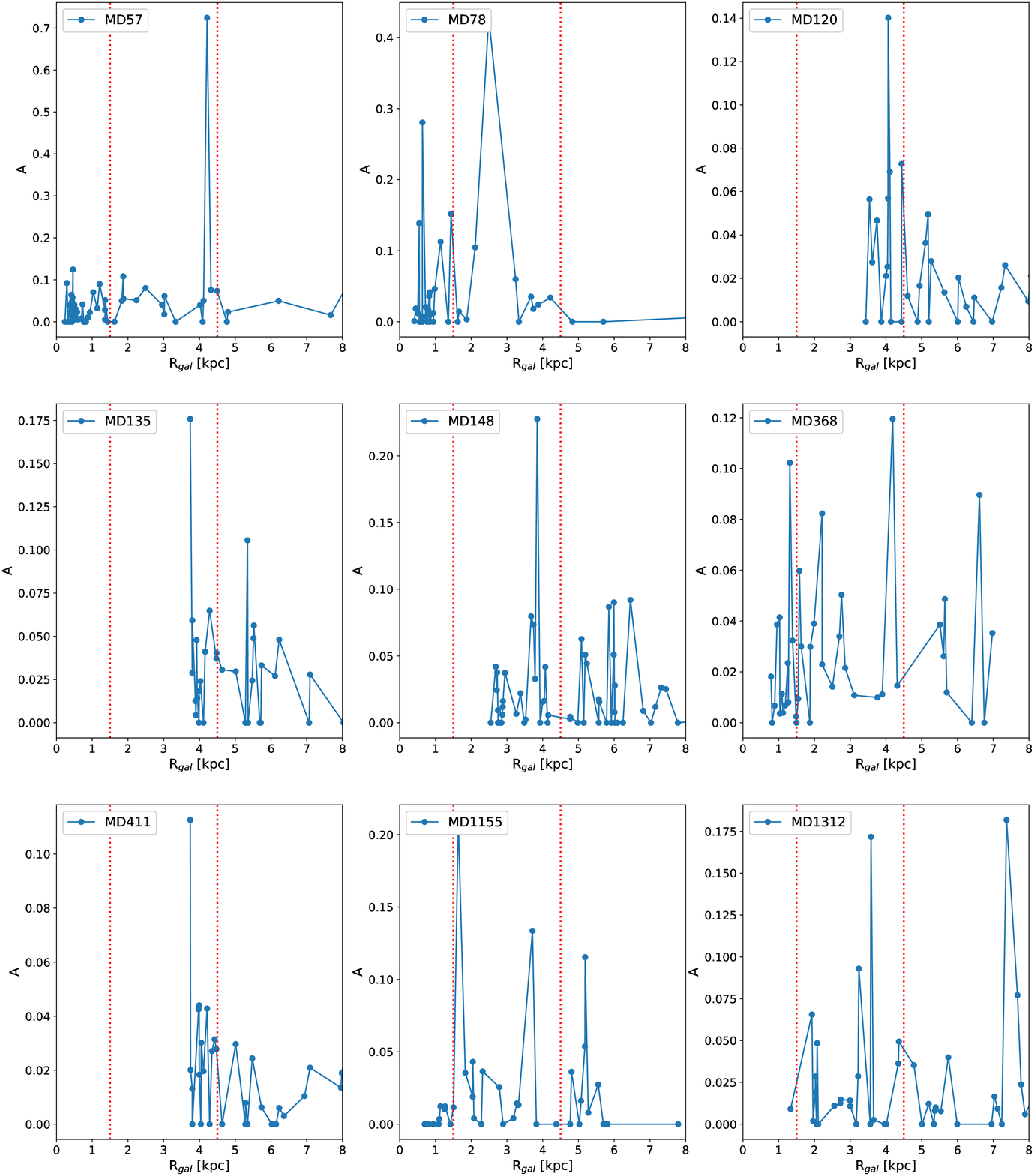}
    \caption{The radial distribution of the clouds that overlap, also partially, with the area of interest. The fraction of $A$ that contribute to the column is plotted. The sum of the $A$ in the 1.5--4.5 kpc ring dominates over the $A$ of clouds in other regions. The values have been taken from \cite{Miville-Deschenes2016}.}
    \label{fig:sub-clouds}
\end{figure}

The gamma-ray counts maps obtained from Fermi-LAT data $>$ 1 GeV after the subtraction of the 4FGL sources are compared to the Planck dust maps in Fig. \ref{fig:gamma-dust}. A good correspondence is observed between the diffuse gamma-ray emission and the dust opacity map. The column under analysis do not emerge as particularly bright source neither in terms of column density, nor in terms of gamma-ray emissivity, but are rather regions of average diffuse gas density. As observed in \cite{Miville-Deschenes2016}, most of l.o.s.s contains only a few molecular clouds (often less than 2). The peculiarity of the chosen observational direction resides in the fact that most of the gas on the line of sight belongs to the 1.5--4.5 kpc region of the Galaxy. The CO along the line of sight of the selected targets is shown in Fig. \ref{fig:los-clouds}. An approximate indication of the galactocentric distance, $R$, corresponding to the given radial velocity, $v$, is derived with the kinematic distance method:

\begin{equation}
R =R_{\odot}V(R)\sin(l)\cos(b) \frac{1}{V_\odot + \frac{v}{\sin(l)\cos(b) }}    
\end{equation}

where $R_\odot=8$ kpc and $V_\odot$=233 km/s are the solar galactocentric distance and rotation velocity. A rotation curve, $V(R)$ as the one given in \cite{Reid2014} is assumed. The edges of the 1.5--4.5 kpc ring, are indicated in the figure. Note that this division is not precise, especially at low galacto-centric distances, because of the uncertainties on the rotation curve and at low longitudes, because of the argument explained in the main text. For this reason, we extended the study on the gas along the selected l.o.s. by using clouds. The galactocentric-distribution of the clouds that fall in the selected regions are shown in Fig. \ref{fig:sub-clouds}. The portion of column density (calculated in terms of A) of each cloud is plotted. The fraction of A that belongs to each cloud is calculated from the number of pixels that fall in the area: for example if only 2 pixels out of 10 overlap the area of interest, only the 20\% of the mass is considered to calculate A. The blue bullets in the plot indicate the center of each cloud.

\bibliography{My_Collection}

\end{document}